\documentclass[a4paper,fleqn,usenatbib,useAMS,letters]{mnras}

\usepackage{amsmath}
\usepackage{graphicx}
\usepackage{times}
\usepackage{enumerate}
\usepackage{enumitem}

\usepackage[T1]{fontenc}
\usepackage{ae,aecompl}

\title[Evidence for Coherent Curvature radiation in J1645$-$0317]{Evidence for Coherent Curvature Radiation in PSR J1645$-$0317 with Disordered 
Distribution of Polarization Position Angle}
\author[Mitra, Melikidze, Basu]{Dipanjan Mitra$^{1,2}$,
George I. Melikidze$^{2,3}$, Rahul Basu$^{2}$\\
$^{1}$ National Centre for Radio Astrophysics, Tata Institute of Fundamental Research, Pune 411007, India \\
$^{2}$ Janusz Gil Institute of Astronomy, University of Zielona G\'ora, ul. Szafrana 2, 65-516 Zielona G\'ora, Poland \\
$^{3}$ Evgeni Kharadze Georgian National Astrophysical Observatory, 0301, Abastumani, Georgia  \\
}

\newcounter{attnctr} 

\usepackage[normalem]{ulem}
\begin{document}
%\date{Accepted\ldots Received\ldots ; in original form\ldots}

%\pagerange{\pageref{firstpage}--\pageref{lastpage}} \pubyear{2017}

\maketitle

\label{firstpage}

\begin{abstract}
The diverse polarization properties in pulsars are in conflict with
applying a unique emission mechanism to the population. The
polarization position angle (PPA) traverse in most pulsars shows a
S-shaped curve that can be interpreted using the rotating vector model
(RVM) as the radio emission being directed either parallel or
perpendicular to the divergent magnetic field lines and argues for a
coherent curvature radiation mechanism from charge bunches in a
strongly magnetized pair plasma. However, in a subset of pulsars the
radio emission is significantly depolarized and the PPA shows a
complex pattern which cannot be explained using RVM. We propose that
even in such cases the highly polarized time samples in the single
pulses should follow the RVM with possibly two parallel tracks
separated by 90\degr. We have investigated PSR J1645$-$0317, with
complex PPA traverse, and demonstrated for the first time that
considering only the highly polarized time samples in the single
pulses, the PPA distribution clearly follows the RVM. We conclude that
this strongly favour the coherent curvature radiation mechanism to be
universally applicable in the pulsar population.
\end{abstract}


\section{Introduction} \label{sec:intro}
Radio pulsar polarization features provide insights into the
properties of relativistic pair plasma in ultra-strong magnetic
fields. However, pulsar polarization properties are still poorly
understood. The linearly polarized radio emission from pulsars
exhibits a polarization position angle (PPA) across the pulsar
profile, which usually follows a characteristic S-shaped curve in
normal pulsars (period $P>0.1$ s) and can be explained using the
rotating vector model \citep[][hereafter RVM]{1969ApL.....3..225R},
where the PPA traces the change in the dipolar magnetic field line
planes. The PPA behaviour shows significant variation in the pulsar
population and can be broadly divided into three categories
\citep{2016ApJ...833...28M} The first category, which we term as
$R_1$, appears in more energetic pulsars, with spin-down energy loss
($\dot{E}$) in excess of 10$^{34}$ erg~s$^{-1}$. The average profile
has linear polarization greater than 80 percent, and the PPA traverse
resembles one single S-shaped curve following the RVM. The second
group, $R_2$, has intermediate energies $10^{32} < \dot{E} < 10^{34}$
erg~s$^{-1}$, and usually has much lower average linear polarization
at around 20 percent. The PPA shows complex pattern both in the
average profile as well as the single pulse distribution that do not
conform to RVM. The final category, $R_3$, is seen in low energetic
pulsars with $\dot{E} < 10^{32}$ erg~s$^{-1}$, and has intermediate
levels of linear polarization around 30-40 percent. The average PPA
traverse can show both a S-shaped curve as well as a complex pattern,
however, the single pulse PPA distribution clearly shows the presence
of two orthogonal S-shaped tracks with 90\degr~separation.

The RVM in radio pulsars suggests both synchrotron and curvature
radiation as possible mechanisms. Synchrotron emission has been deemed
unsuitable since the extremely ordered magnetic fields required, can
only occur well inside the light cylinder where the synchrotron
frequency is outside the radio frequency range
\citep{1969PASA....1..254R}. The vacuum curvature radiation is another
possible emission mechanism that requires the electric field vectors
to lie in the plane of the charge trajectory. Detailed measurements
have shown the orientation of the emerging electric vectors in pulsars
to be perpendicular to the dipolar magnetic field line plane
\citep[see e.g.][]{2001ApJ...549.1111L}, thereby ruling out the vacuum
curvature radiation as the likely emission mechanism. Hence, an
alternative model like coherent curvature radiation, excited in
electron-positron plasma by charge bunches, is preferred. This was
first suggested by \citet{1975ApJ...196...51R}, where sparking
discharges at the vacuum gap gave rise to a non-stationary plasma flow
and subsequently the development of two stream instability of Langmuir
mode could result in the formation of charge bunches. There were
initial concerns regarding the stability of the charge bunches
\citep{1995JApA...16..137M}, however, subsequent studies have shown
that when modulational instability of Langmuir mode is considered,
solutions for stable charge solitons can be obtained that resemble
charge bunches \citep{1980Afz....16..161M,2000ApJ...544.1081M,
  2004ApJ...600..872G,2018MNRAS.480.4526L,2022MNRAS.516.3715R}. Other
ways of obtaining charge bunches include two-stream instability of
counter-streaming pairs
\citep{2020ApJ...901L..13Y,2021A&A...649A.145M}, electrostatic
oscillations during sparking events \citep{2020MNRAS.494.2385K,
  2021ApJ...908..149C} or recent studies of global magnetospheric
simulations in the polar cap region \citep{2022arXiv220911362B}.

The eigen-modes of the dielectric response tensor in a strongly
magnetized electron-positron plasma consist of two sub-luminal,
linearly polarized modes \citep[see e.g.][]{1986ApJ...302..120A} : the
purely electromagnetic, transverse, $t$-mode, and the
longitudinal-transverse, $lt_1$-mode, \citep[using terminology
  from][]{2003PhRvE..67b6407S}. The electric field vector of the
$t$-mode, $\vec E_t$, is perpendicular to the plane of the
wave-vector, $\vec k$, and local magnetic field, $\vec B$, hence
${\vec E_t}\perp{\vec k}$. The electric field vector of the
$lt_1$-mode, $\vec E_{lt}$, is parallel to the plane containing $\vec
k$ and $\vec B$, and as a result ${\vec E_{lt}}$ has a component along
${\vec k}$. If ${\vec k}\parallel{\vec B}$ then the two modes
coincide. The coherent curvature emission can only excite these two
eigen-modes of the surrounding plasma \citep{2004ApJ...600..872G}. Due
to differences in their refractive indices the two modes split-up
during propagation in the plasma. The $lt_1$-mode, which is seven
times stronger, is ducted along the magnetic field lines and is unable
to escape from a homogeneous plasma, contrary to the $t$-mode which
can freely propagate in it \citep{2004ApJ...600..872G,
  2014ApJ...794..105M,2022MNRAS.512.3589R}. But the magnetospheric
plasma, comprising of columns with variable density, is not expected
to be homogeneous
\citep{2003A&A...407..315G,2021ApJ...919L...4C}. Hence, a part of the
$lt_1$-mode is expected to escape from the dense inhomogeneous plasma
along with the $t$-mode and propagate as electromagnetic waves, namely
the ordinary, O-mode, and extraordinary, X-mode, polarized parallel
and perpendicular to the magnetic field line planes, respectively
\citep[see][]{2014ApJ...794..105M,2021MNRAS.500.4549M}.

The radio emission from pulsars originates due to a large number of
charged bunches emitting coherently, with the observed polarization
feature obtained from incoherent addition of orthogonally polarized
waves from these bunches. If one of the modes (most likely the
$lt_1$-mode) is significantly damped, then S-shaped PPA traverse
similar to the $R_1$ category of pulsars will be observed. When two
orthogonal modes with random amplitudes are incoherently added, one
expects the observed polarization behaviour to exhibit large scatter
with complex PPA as seen in the $R_2$ category of pulsars. Under
certain conditions both the orthogonal modes emerge from the plasma
and two PPA tracks across the pulse window, separated by 90\degr, is
observed, and corresponds to the $R_3$ category. The above model also
clearly predicts that during instances when the observed signal is
nearly 100\% linearly polarized, i.e. no significant effect of
depolarization takes place due to incoherent mixing of the orthogonal
modes, the PPA should follow the RVM.  \citet{2009ApJ...696L.141M} and
\citet{2014ApJ...794..105M} showed that the highly linearly polarized
single pulses from several pulsars, in the $R_1$ and $R_3$ category,
have PPA which can be modelled by the RVM. This was considered as
conclusive observational evidence for the coherent curvature radiation
as the radio emission mechanism for highly linearly polarized signals
in these pulsars. While it is possible that the emission mechanism for
weakly polarized pulses are also due to coherent curvature radiation,
nonetheless, as discussed in
\citet{2016JPlPh..82c6302E,2017JPhCS.932a2011M, 2021MNRAS.500.4530M},
the conclusion about the emission mechanism might not be so
straightforward and these pulses can have a different underlying
emission mechanism. In this work we have analyzed the highly linearly
polarized signals from the bright pulsar, PSR J1645$-$0317, which
belongs to the $R_2$ category, to test the applicability of the
coherent curvature radiation mechanism in it.

\section{Observation \& Analysis} \label{sec:obs}

\begin{figure*}
\begin{center}
\begin{tabular}{cc}
\mbox{\includegraphics[width=9cm,height=5.0cm,angle=-00]{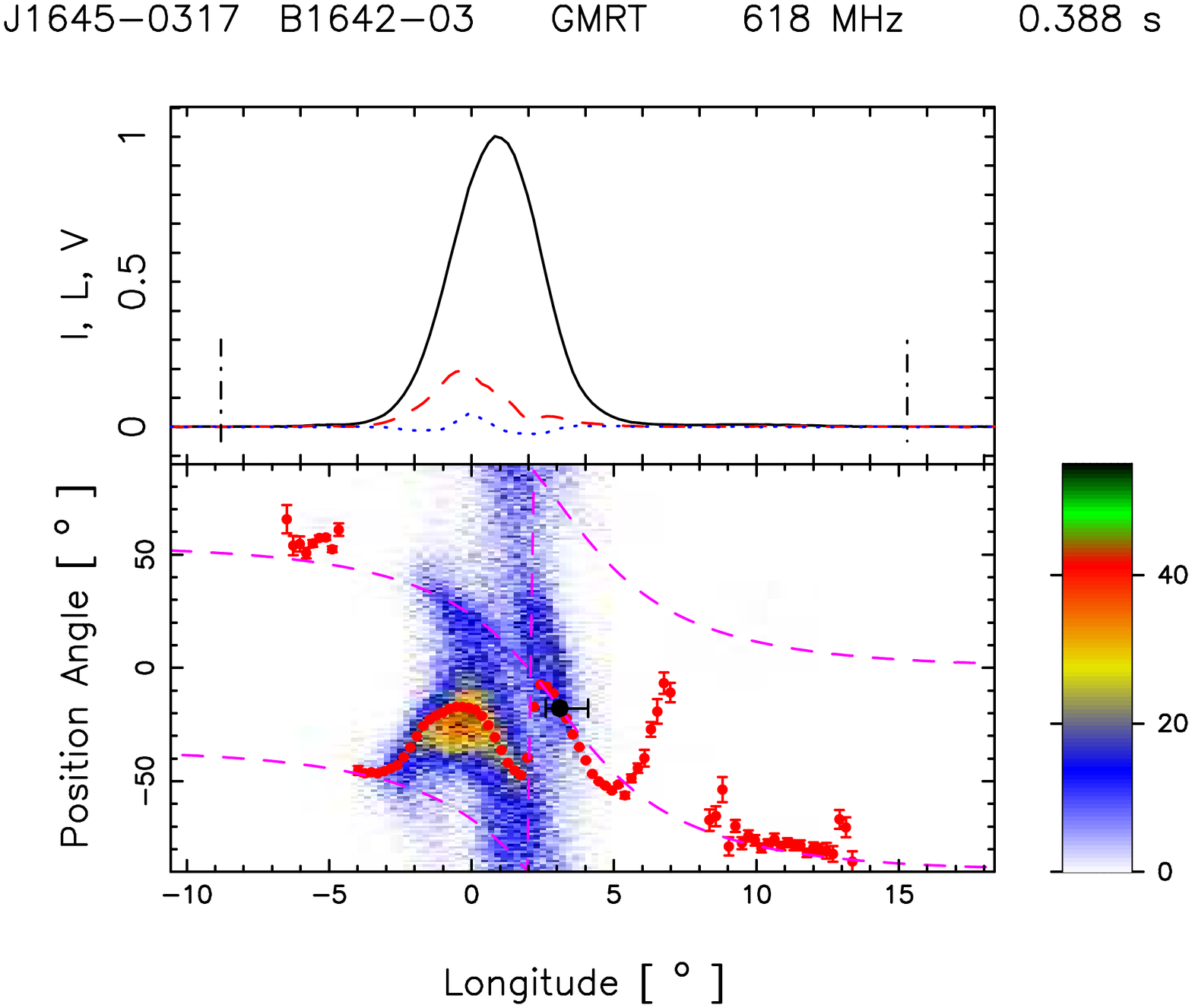}} &
\mbox{\includegraphics[width=9cm,height=5.0cm,angle=-00]{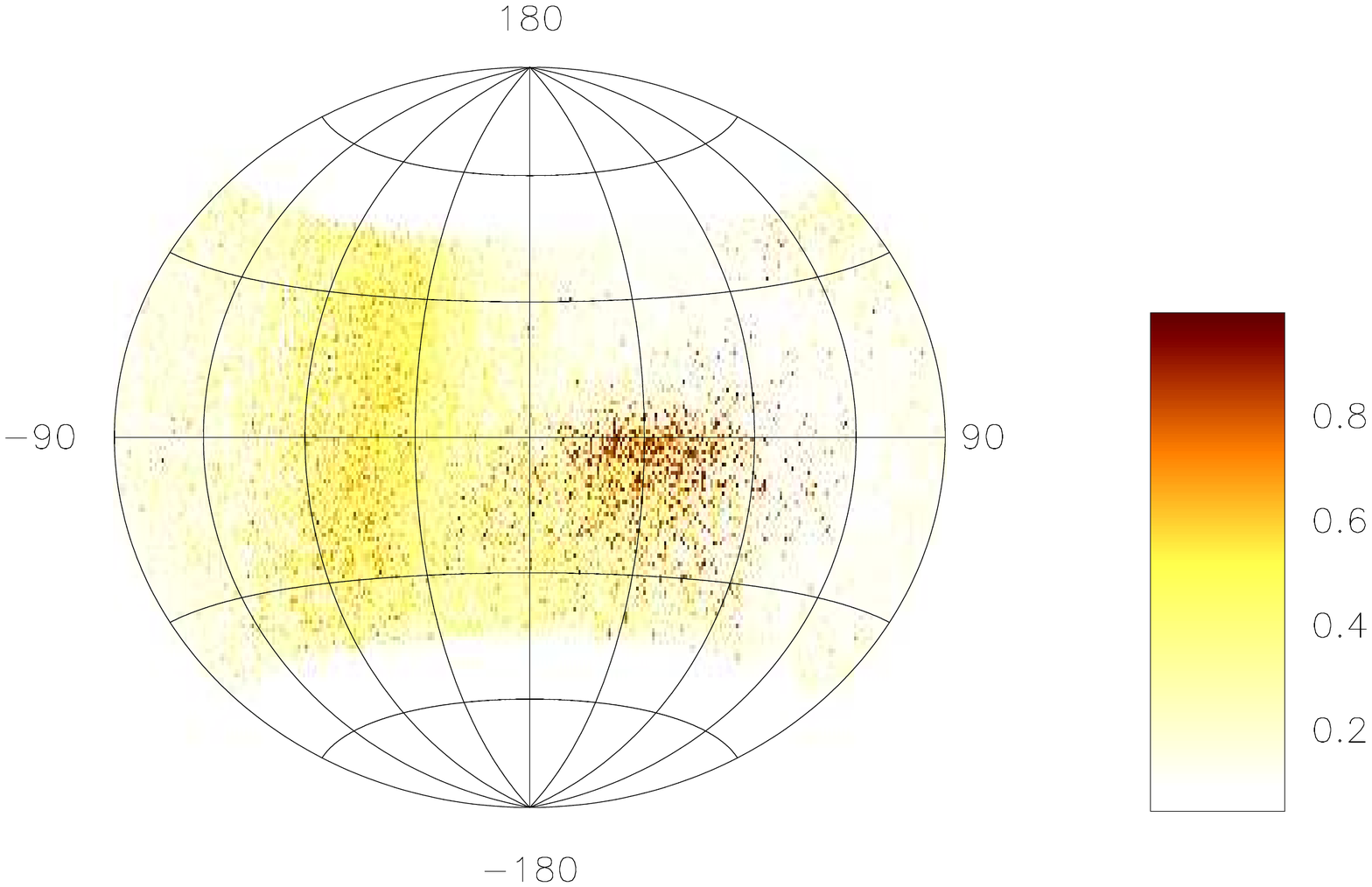}} \\
	(a) & (b) \\
\mbox{\includegraphics[width=9cm,height=5.0cm,angle=-00]{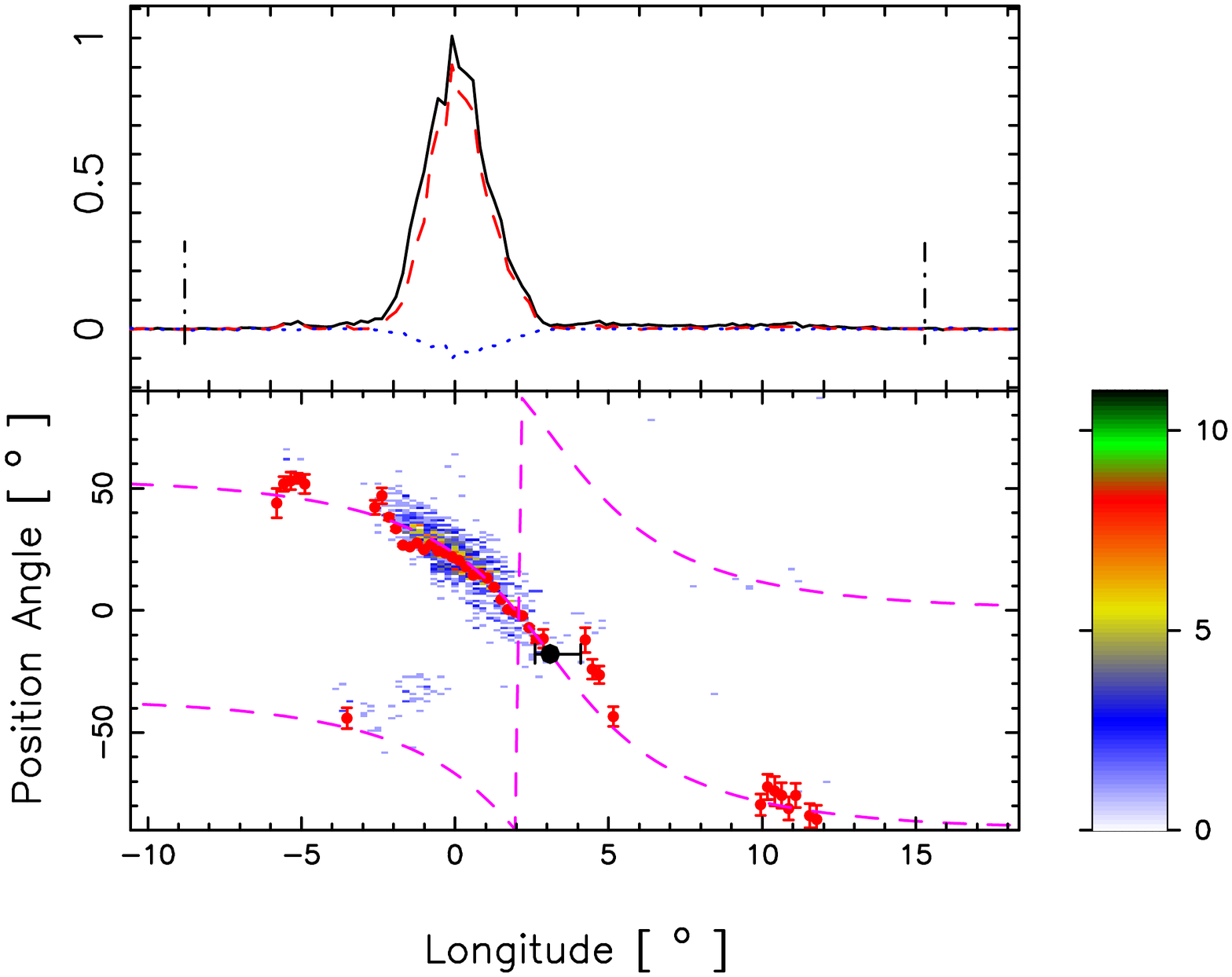}} &
\mbox{\includegraphics[width=9cm,height=5.0cm,angle=-00]{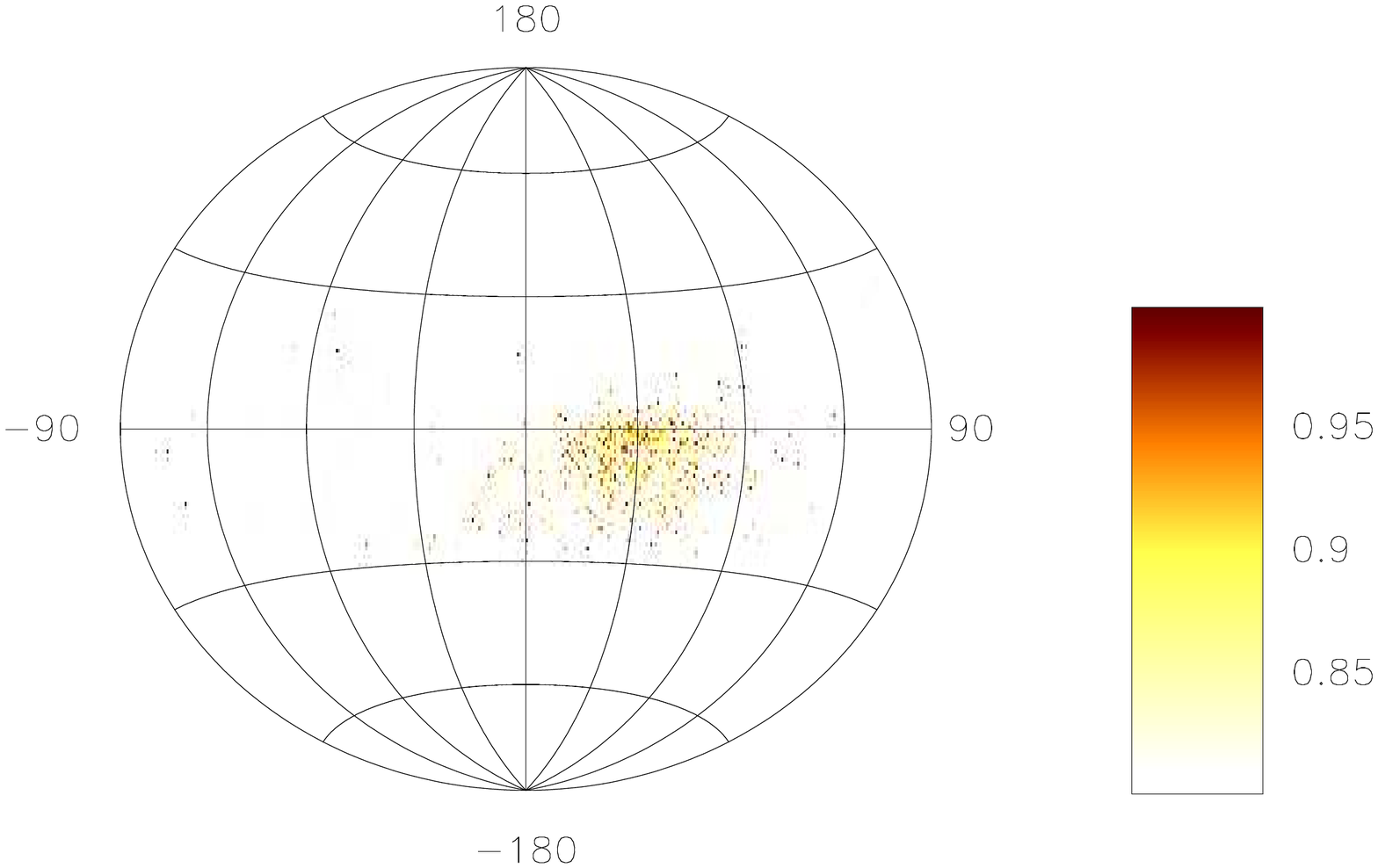}} \\
	(c) & (d) \\
\mbox{\includegraphics[width=9cm,height=5.0cm,angle=-00]{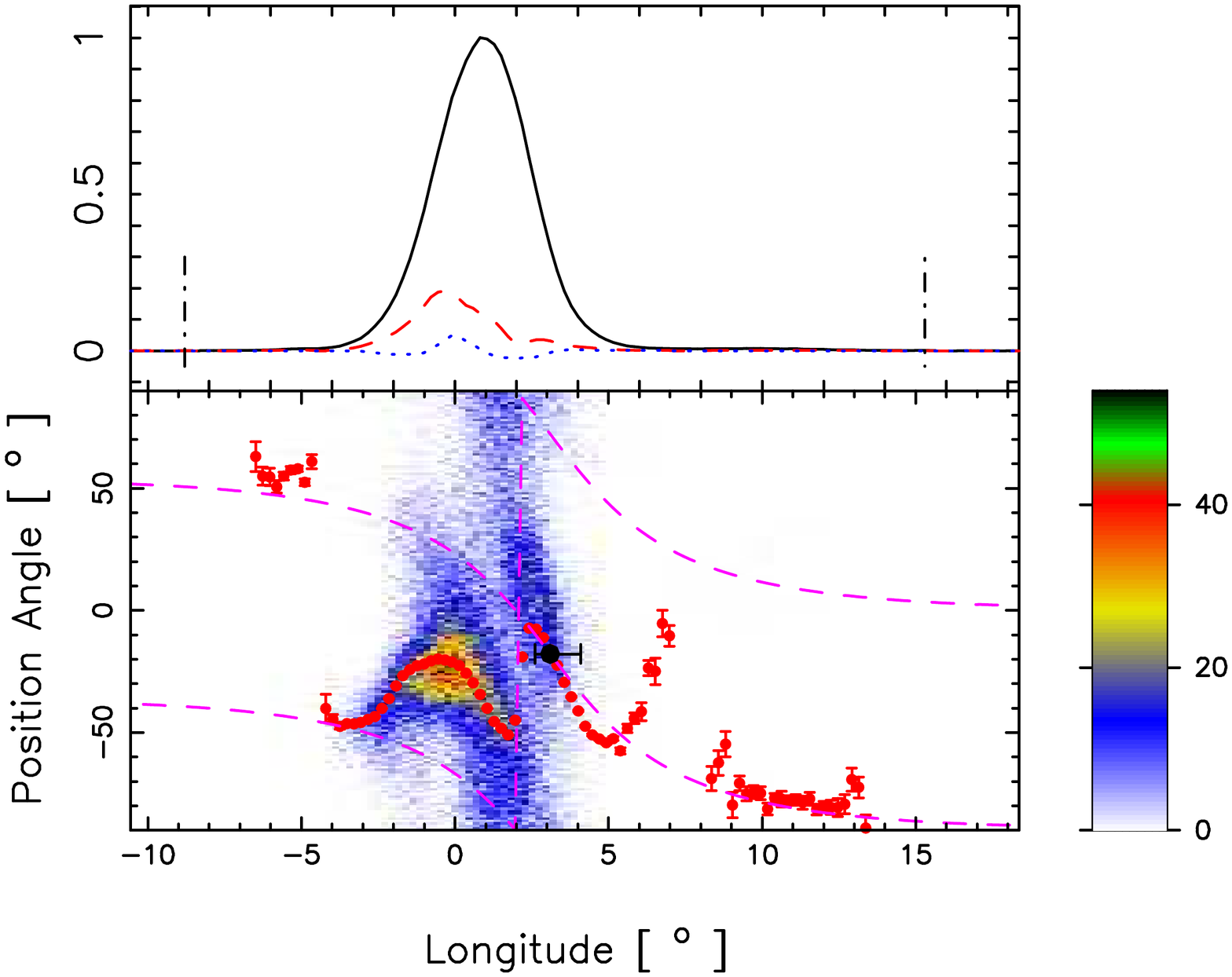}} &
\mbox{\includegraphics[width=9cm,height=5.0cm,angle=-00]{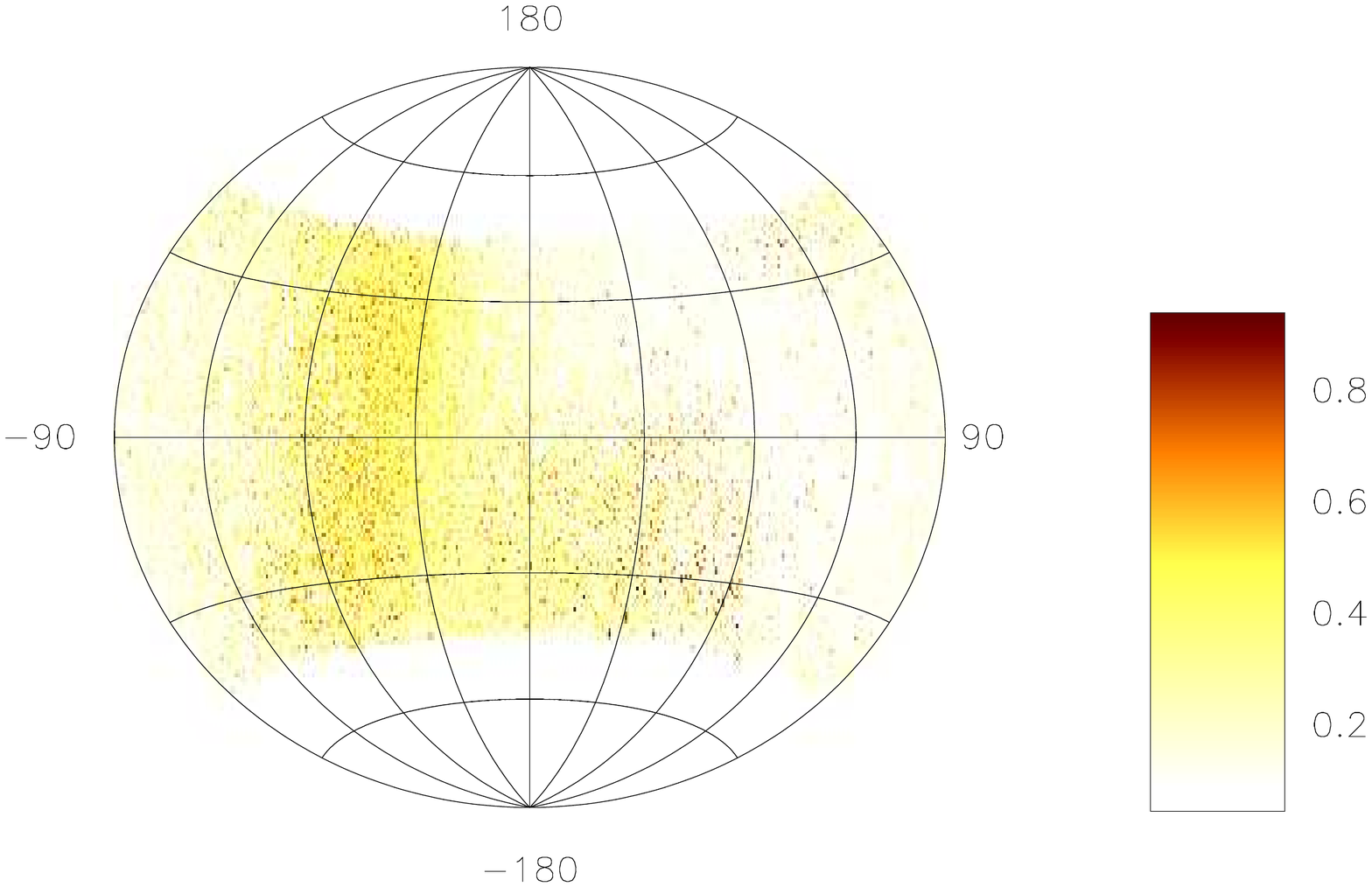}} \\
	(e) & (f) \\
\hspace{40px}
\end{tabular}
\caption{The polarization behaviour of PSR J1645$-$0317~in the average
  profile as well as single pulse distributions. The top panels (a)
  and (b) correspond to the entire observing duration, the middle
  panels (c) and (d) show the time samples with more than 80 percent
  linear polarization, and the bottom panels (e) and (f) correspond to
  time samples with less than 70 percent polarization.  The upper
  windows in left panels (a), (c) and (e) show the average profile
  with total intensity (Stokes I; solid curve), total linear
  polarization (dashed red), circular polarization (Stokes V ; dotted
  blue), while in the lower panels the single pulse polarization
  position angle (PPA) distribution is shown in colour scale along
  with the average PPA as red error bars. In panels (c) and (e) we
  have also shown the rotating vector model (RVM) fits to the PPA
  (magenta dashed line) which is plotted twice with 90\degr~separation
  to show the two polarization modes. The geometric parameters used
  for RVM fits are $\alpha=165\degr$, $\beta=-0.9\degr$,
  $\phi_{\circ}=3.1\degr$ and $\psi_{\circ}=-17.8\degr$ (see text for
  definitions). The right panels (b), (d) and (f) shows the
  Hammer-Aitoff projection of the polarized time samples in each case
  and the colour scheme shows the fractional polarized power.}
\label{f1}
\end{center}
\end{figure*}

We have used archival observations of PSR J1645$-$0317 at 618 MHz
observed in Meterwavelength Single-pulse Polarimetric Emission Survey
\citep[MSPES,][]{2016ApJ...833...28M} conducted with the Giant
Meterwave Radio Telescope to investigate the single pulse polarization
behaviour. PSR J1645$-$0317 is classified as a normal pulsar with
$P=0.388$ seconds, characteristic age 3.45 Myr and
$\dot{E}=1.21\times10^{33}$ ergs~s$^{-1}$. The pulsar is relatively
bright with average flux density at 618 MHz, $S_{618}$ = 291.2$\pm$8.1
mJy \citep{2021ApJ...917...48B}. PSR J1645-0317 was observed on 7
March, 2014, continuously for 14 minutes, resulting in 2154 single
pulses.  The minimum time resolution of polarization observation from
GMRT was 0.246 milliseconds. This resulted in 1577 phase-bins across
the pulsar period and 103 phase-bins across the pulse window. The
baseline noise levels in the average profile was estimated to be 2.2
mJy, which ensured that the single pulses were detected with high
sensitivity for polarization studies. The standard polarization
calibrator, PSR B1929+10, was observed every two hours to account for
variations, if any, in the polarization response of the telescope
receiver system and obtain the four stokes parameters
($I,Q,U,V$). Using the definition in MSPES we obtain the average
linear polarization, $L(\phi)$, circular polarization, $V(\phi)$, and
PPA, $\psi_(\phi)$, for different phase longitudes, $\phi$, across the
pulse profile.

The polarization behaviour is summarized in Fig.\ref{f1} where we have
estimated three different realizations, the entire single pulse
sequence in top panels (a) and (b), the highly polarized signals with
linear polarization level above 80 percent in middle panels (c) and
(d), and time sample with polarization less than 70 percent in bottom
panels (e) and (f). The top window in left panels (a), (c) and (e)
shows the average behaviour, including the total intensity profile
(black line) the average linear polarization (red dashed line) and
average circular polarization (blue dotted line). The bottom window in
each case shows the PPA distribution of the time samples corresponding
to the phase-bins across the pulse window. The PPA distribution
considered polarized time samples in the single pulses with linear
polarization above three times the baseline noise levels, while the
average PPA (red error bars) used similarly significant signals from
the average profile.  The colour scheme for the PPA plots correspond
to the number of time samples in the distribution. The right panels
(b), (d) and (f) presents the stokes parameters as the Hammer-Aitoff
projection of the Poincar\'e sphere for the different times samples,
with the colour scheme showing the fractional polarization level,
$\sqrt{L^2 + V^2}/I$. Note that out selection criteria yielded 36678
PPA time samples in panels (a) and (b), 1057 samples in panels (c) and
(d), and 34604 samples in panels (e) and (f).

\section{Results}

\subsection{Profile Classification and RVM estimate}
The average profile of PSR J1645$-$0317 shows one prominent component
which resembles the core emission in the core-cone model of the pulsar
emission beam \citep{1993ApJ...405..285R}. The conal component is
weaker at meterwavelengths but much more prominent at higher
frequencies and the pulsar has been classified as a Triple
profile. \citet{2022ApJ...927..208B} showed that the central component
has a steeper spectra compared to the surrounding conal emission with
relative spectral index $\Delta=-1.39\pm0.05$, a characteristic
feature of the core emission.

In RVM, the PPA, $\psi$, at pulse phase, $\phi$, is related to the
angle between the rotation axis and magnetic axis, $\alpha$, and the
angle between the magnetic axis and the line of sight (LOS), $\beta$,
as :
\begin{equation}
\psi = \psi_{\circ} + \tan^{-1} \left(\frac{\sin{\alpha}\sin{(\phi-\phi_{\circ})}}
{\sin{(\alpha + \beta)}\cos{\alpha} - \sin{\alpha}\cos{(\alpha+\beta)}\cos{(\phi-\phi_{\circ})}}\right).
\label{req1}
\end{equation}
Here $\psi_{\circ}$ and $\phi_{\circ}$ are the phase offsets for the
PPA and the pulse phase, respectively. The core emission lies at the
center of the emission beam and entails the LOS traverse to be close
to the magnetic axis. As a result the steepest gradient (SG) point of
the PPA traverse $\mid d\psi/d\phi \mid_{max} =
\sin(\alpha)/\sin(\beta) > 10$ in core pulsars.

Earlier it has not been possible to estimate the RVM fits in $R_2$
category pulsars, like PSR J1645$-$0317, due to their complex PPA
traverses. However, we have now shown in Fig.\ref{f1}(c) that by
considering only highly polarized signals the resulting PPA traverse
closely resemble the RVM, with a narrow spread and a dominating
polarization mode. The Hammer-Aitoff projection in Fig.\ref{f1}(d)
also shows these time samples to be distributed around the equator,
suggesting high levels of linear polarization. The density is highest
at the center of the tracks, and has time samples with $<10$ percent
circularly polarized power. Fig.\ref{f1}(c) also shows the RVM fits of
eq.(\ref{req1}), to the modified PPA and we have estimated $\mid
d\psi/d\phi \mid_{max} \sim 16.5\pm2$. The high SG value of the PPA
traverse provides further evidence for the central component to be
core emission.

\subsection{Orthogonal and Non-Orthogonal Polarization Modes}
Fig.\ref{f1}(c) shows in addition to the dominant PPA distribution a
second clustering around $(\phi,\psi)=(-2\degr,-40\degr)$, which
raises the possibility of detecting the orthogonal mode. However, on
closer inspection we found this grouping to be separated by around
60\degr~with respect to the dominant mode and not 90\degr as would be
expected for an orthogonal mode. Additionally, these points are
elliptically polarized with more that 20\% circular polarization. We
conclude these points do not correspond to the orthogonal mode in
general but is reminiscent of the polarization mixing. It is likely
that the occurrence of orthogonal polarization mode is rare in this
pulsar and longer observations with higher sensitivity would be
required to clearly identify them.

Fig.\ref{f1}(e) shows the PPA distribution of the depolarized signals
in the single pulses, along with the RVM fits obtained from highly
polarized signals in Fig.\ref{f1}(c). Most of the distribution is
confined between the two RVM curves suggesting that the depolarized
signals can have arbitrary PPA values.

\subsection{Radio Emission Height}
Using effects of aberration and retardation
\citep{1991ApJ...370..643B}, the radio emission height can be
estimated as $h_E = c P \Delta \phi/1440$ km, where $c$ is speed of
light in km~s$^{-1}$ and $P$ is in seconds. Here,
$\Delta\phi=\phi_{\circ}-\phi_c$, $\phi_c$ being the center of the
profile and $\phi_{\circ}$ is the location of the SG point of the PPA,
obtained from RVM fits (see eq.\ref{req1}). In well defined profiles,
$\phi_c$ is obtained from the leading and trailing edges, which
correspond to the last open dipolar field lines
\citep{2004A&A...421..215M}. However, in PSR J1645$-$0317 the outer
conal components are weak at 618 MHz and leads to erroneous
detections. On the other hand the identification of the central
component as core, allowed us to use its peak location as an alternate
estimate of $\phi_c$. We found $\Delta\phi=3.1\pm0.5\degr$ and the
corresponding core emission height $h_E=250\pm40$ km. Since RVM is
independent of emission height, the emission across the profile is
expected to arise from similar heights $\sim250$ km, consistent with
the height obtained for normal pulsars
\citep{1997A&A...324..981V,2008MNRAS.391.1210W,2016MNRAS.460.3063M}.

\section{Discussion}
It is difficult to reconcile the disordered PPA distributions in the
$R_2$ category of pulsars as they do not seem to follow the RVM. Our
analysis of the highly polarized time samples in PSR J1645$-$0317
shows the PPA to exhibit a S-shaped curve consistent with the RVM
fits. In addition, we were able to estimate the radio emission height
to be well below 10 percent of the light cylinder radius, which is
typical for normal pulsars. These results clearly demonstrate that
radio emission mechanism for highly polarized time samples is
identical in the pulsar population including the $R_2$ category that
exhibit disordered PPA distributions in the average profile. The
single pulse polarization studies of \citet{2009ApJ...696L.141M}
showed that the only viable emission mechanism that can explain the
RVM behaviour of the PPA traverse of highly polarized pulses is
coherent curvature radiation from charge bunches, and hence this
mechanism may be universally applicable to the entire normal pulsar
population.

 The theory of coherent curvature radiation from charge bunches has
 been developed for the model of plasma being generated from sparking
 discharges, where charge bunches are formed in dense,
 spark-associated, pair plasma clouds
 \citep{2000ApJ...544.1081M,2004ApJ...600..872G}. The coherent
 curvature radiation in PSR J1645$-$0317 gives estimate of the Lorentz
 factor, $\gamma$, of charge bunches as $\gamma \sim (2 \nu_c
 \rho_c/3c)^{1/3} \sim 100$, where the characteristic frequency is the
 observing frequency, $\nu_c \sim 600$ MHz, $c$ is the velocity of
 light, $\rho_c \sim 10^8$ cm the radius of curvature at the radio
 emission height $h_E \sim 250$ km. Further, the observed emission is
 generated by a large number of charge bunches, even at the smallest
 time resolution of observation.

The characteristic plasma frequency in the clouds can be estimated as
$\nu_{\circ} \sim 2 \times 10^{-5} \kappa^{0.5} \sqrt{\gamma}
(\dot{P}_{-15}/P^7)^{0.25} (2\pi h_E/c P)^{-1.5}$ GHz. In PSR
J1645$-$0317, $P=0.387$ s, $\dot{P}_{-15} = 1.78$ and using
multiplicity $\kappa = 10^5$ and $h_E=250$ km, we obtain $\nu_{\circ}
\sim 245$ GHz which is significantly larger than 600 MHz. The charge
bunches excite the linearly polarized $t$-mode and $lt_1$-mode, which
separates in the plasma cloud and eventually escapes as
electromagnetic waves from the edge of the plasma cloud boundary,
where large plasma density gradient exist, into the low density inter
cloud region.  Due to the presence of large number of charge bunches,
the resultant linear polarization is likely to be an incoherent
addition of the X-mode and O-mode, and provides a possible explanation
for the significant depolarization and the disordered PPA distribution
observed in PSR J1645$-$0317 (see Fig.\ref{f1}, lower panels). It is
statistically likely that at certain instances one of the modes is
suppressed and the corresponding emission will show high levels of
polarization with PPA showing RVM nature (see Fig.\ref{f1}, middle
panels).

Generally the plasma waves corresponding to the X-mode ($t$-mode) and
O-mode ($lt_1$-mode) are polarized perpendicular and parallel to the
plane of the local magnetic field and the wave-vector,
respectively. If we presume the existence of an isolated source
emitting either the $t$-mode or $lt_1$-mode excited by some unknown
emission mechanism, then depending on the mutual orientation of the
wave-vector and magnetic field one can indeed observe the linearly
polarized emission and S-shaped nature of the PPA. On the other hand
if there exist a large number of elementary emitters instead of a
single one, then incoherent addition of the $t$-mode (or $lt_1$-mode)
will result in depolarization and disordered PPA
distribution. However, the observations reported here clearly show
that even in the smallest time resolution bin, the PPA corresponding
to the highly linearly polarized emission traces the plane of the
curved dipole magnetic field line plane in accordance with the
RVM. This leaves us with the only possibility that all elementary
emitters that radiate the $t$-mode (or $lt_1$-mode), are polarized in
a particular way, i.e.  perpendicular (or parallel) to the local
curved magnetic field line plane. The only known mechanism that makes
this possible is curvature radiation.

The coherent curvature radiation mechanism can explain the overall
observed linear polarization features in principle by an incoherent
superposition of the $t$-mode and $lt_1$-mode. However, only in the
case of highly linearly polarized signals, the RVM fits to the PPA
provide conclusive evidence for the presence of curvature
radiation. In PSR J1645$-$0317 a majority of the time samples do not
follow the RVM and have polarization fraction below 70\%. In these
cases we cannot exclude the possibility that they might be formed due
to an alternative emission mechanism operating alongside the curvature
radiation mechanism.

\section{Summary}
In this work we have demonstrated that the PPA corresponding to
signals that show high levels of linear polarization, closely follow
the RVM, even in pulsars that have complex PPA traverse. This clearly
suggests that the highly linearly polarized radio emission from
pulsars are akin to the properties of the eigen-modes corresponding to
the $t$-mode and the $lt_1$-mode of electron-positron plasma in strong
magnetic field, excited by coherent curvature radiation from charge
bunches. We propose that these modes can escape as perpendicularly
polarized extraordinary waves and parallelly polarized ordinary waves,
with respect to magnetic field line planes, in an inhomogeneous plasma
with plasma density gradients. The detailed model of how the $t$-mode
and the $lt_1$-mode escape from the boundary between plasma clouds
will be explored in a future work.

\section*{acknowledgments}
We thank the anonymous referee for helpful comments that improved the
quality of the manuscript.  We thank the staff of the GMRT that made
these observations possible.  D.M. acknowledges the support of the
Department of Atomic Energy, Government of India, under project
No. 12-R\&D-TFR-5.02-0700. D.M. acknowledges funding from the grant
``Indo-French Centre for the Promotion of Advanced Research -
CEFIPRA'' grant IFC/F5904-B/2018. This work was supported by grant
2020/37/B/ST9/02215 from the National Science Center, Poland.

\section*{Data availability}
The data underlying this article will be shared on reasonable request
to the corresponding author.

\bibliographystyle{mn2e}
\bibliography{References}
\end{document}